\documentclass{aa}
\usepackage{graphicx,epsfig}

\begin{document}
   \title{Geometry and Kinematics in the Central Broad-Line
          Region of a Seyfert\,1 Galaxy}

   \author{W. Kollatschny
          \inst{1,2}\fnmsep
          \thanks{Based on observations obtained with the Hobby-Eberly
               Telescope, which is a joint project of the University
               of Texas at Austin, the Pennsylvania State
           University, Stanford University, Ludwig-Maximilians-Universit\"at
            M\"unchen, and Georg-August-Universit\"at G\"ottingen.}
         \
         , K. Bischoff\inst{1}
          }

   \offprints{W. Kollatschny}

   \institute{Universit\"{a}ts-Sternwarte G\"{o}ttingen,
              Geismarlandstra{\ss }e 11, D-37083 G\"{o}ttingen, Germany\\
              \email{wkollat@uni-sw.gwdg.de}
         \and
             Department of Astronomy and McDonald Observatory,
             University of Texas at Austin, Austin, TX 78712, USA}

   \date{Received date, 2002; accepted date, 2002}

   \abstract{
We recorded spectra of the highly variable Seyfert 1
galaxy Mrk\,110 in a variability campaign
with the 9.2m Hobby-Eberly Telescope at McDonald Observatory in order to
study the detailed line profile variations of the broad emission lines.
Here we show that only an AGN model predicting the formation of the broad
H$\beta$ line emission in the wind of an accretion disk 
matches the observed 2-D variability pattern.
Furthermore, we derive an improved mass 
of the central supermassive black hole of
$M = 1.0^{+1.0}_{-0.5}\times 10^{7} M_{\odot}$
from the H$\beta$ velocity-delay map.
   \keywords{Accretion disks --
                Line: profiles --
                Galaxies: Seyfert  --
                Galaxies: individual:  Mrk\,110 --
                Galaxies: nuclei --
                (Galaxies:) quasars: emission lines
               }              
      }
  \authorrunning{W. Kollatschny et al.}
  \titlerunning{Geometry and kinematics in the Central Broad-Line Region}

   \maketitle
%

\section{Introduction}

In the central regions of active galactic nuclei (AGN)
broad emission lines are generated. These emission lines are emitted
in a photoionized gas at distances of light days to light months 
from a central supermassive black hole.
But many details
 of the innermost AGN region
where the broad emission 
lines originate are unknown.
We do not know the geometry
of the line emitting clouds: are they distributed
spherically, in a disk geometry, or biconically?
Different kinematic models of the clouds are conceivable:
radial inflow or outflow motions including accelerated outflow,
turbulent/chaotic velocity fields, 
or (randomly oriented) Keplerian orbits.

The response of variable broad emission lines
to continuum variations can provide 
information on size and geometry of the BLR in AGN.
Especially the compilation of velocity-delay diagrams of variable emission
line profiles is extremely valuable.
But, very homogeneous data sets obtained with high S/N
 are needed for constructing such diagrams.
 The ionizing flux of the
 active galaxies to be analyzed
 should be highly variable.
Furthermore, the spectra have to be acquired over periods of many months
with sampling rates of the order of days.
So far only preliminary results of velocity-delay maps of other galaxies
exist because of insufficient S/N ratio of the data.
Velocity-delay maps have been derived for the Balmer lines in NGC\,5548
(Kollatschny \& Dietrich \cite{kollatschny96}) and NGC\,4593
 (Kollatschny \& Dietrich \cite{kollatschny97}) 
and for the CIV1550 line in NGC\,5548
(Done \& Krolik \cite{done96}) and NGC\,4151 (Ulrich \& Horne \cite{ulrich96}).
The CIV1550 line is often hampered by very strong central absorption
and blended with the HeII1640 line.
Here we present velocity-delay maps of the H$\beta$ line in Mrk\,110.

\section{Observations and data analysis}

We have chosen Mrk 110 as our first target of a thorough variability
campaign with the 9.2m Hobby-Eberly Telescope (HET)
at McDonald Observatory.
From earlier campaigns we knew 
the extreme continuum and line intensity variations in this galaxy
(Bischoff \& Kollatschny \cite{bischoff99}, Peterson et al. \cite{peterson98}).
We took 26 spectra of this 
narrow-line Seyfert 1 galaxy between
1999 November 13 and 2000 May 14. All observations were made under
identical conditions with exactly the same instrumentation at the HET.
The spectra cover the wavelength range from
4200\AA\ to 6900\AA\ with a resolving power of 650 at 5000\AA\,. 
We reduced the data in a homogeneous way
with IRAF reduction packages. In most cases we yielded
a S/N $>$ 100 per pixel in the continuum.
Details of the campaign and reduction procedure are published in
Kollatschny et al. (\cite{kollatschny01}, hereafter Paper\,I).

\section{Results and Discussion}

\subsection{Geometry and kinematics}

In Paper\,I we verified an ionization stratification in the 
broad-line region of Mrk\,110. The integrated
emission line intensities showed different delays
 with respect to continuum variations.
The cross-correlation
of the integrated H$\beta$ light curve with the 
continuum light curve yielded a mean radius of 24.2 light days
of the H$\beta$ emitting region. 

In this paper we discuss H$\beta$ line profile variations of our
Mrk\,110 variability campaign. The H$\beta$ line is the strongest
undisturbed emission line in
the spectra. Furthermore it is arranged
 close to our primary internal flux calibrator [OIII]5007.
The redshifted H$\alpha$ line is heavily contaminated by
 atmospheric absorption. 
The normalized mean H$\beta$ profile of our spectra
is shown in Fig.\ 1 in velocity space.
\begin{figure}
\includegraphics[bb=40 60 400 700,width=55mm,height=85mm,angle=270]
{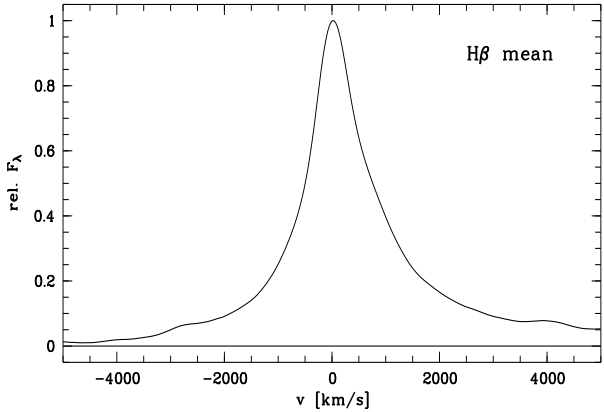}
\caption{Normalized mean H$\beta$ line profile of Mrk~110 in velocity space.}
\end{figure}
We generated light curves from various
H$\beta$ velocity segments ($\Delta v$ = 200 km/s).
Light curves of the optical continuum, of
the H$\beta$ line center as well as blue and red H$\beta$
line wing segments are shown in Fig.\  2.
\begin{figure*}
 \hbox{\includegraphics[bb=40 90 380 700,width=55mm,height=85mm,angle=270]
{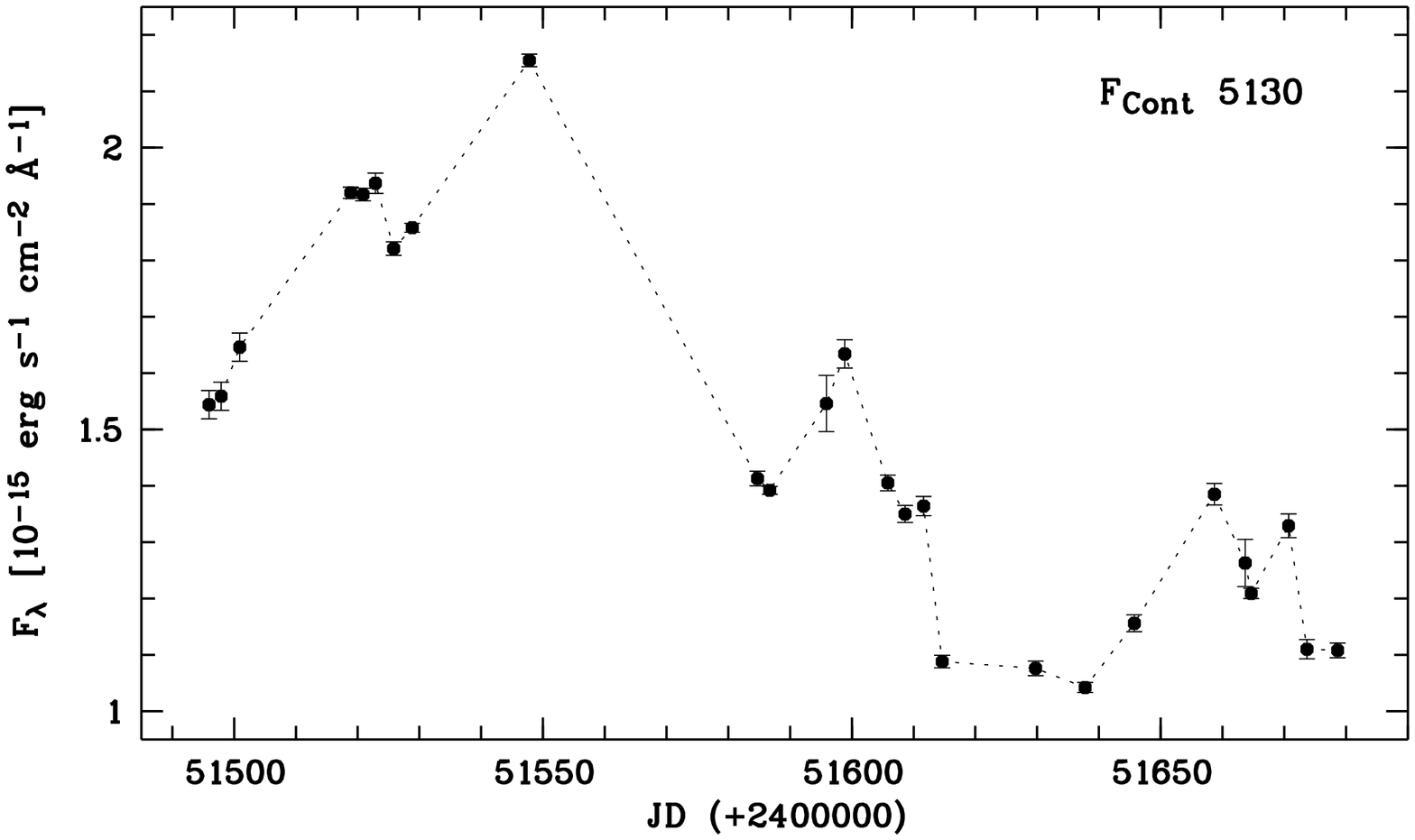}\hspace*{7mm}
       \includegraphics[bb=40 90 380 700,width=55mm,height=85mm,angle=270]
{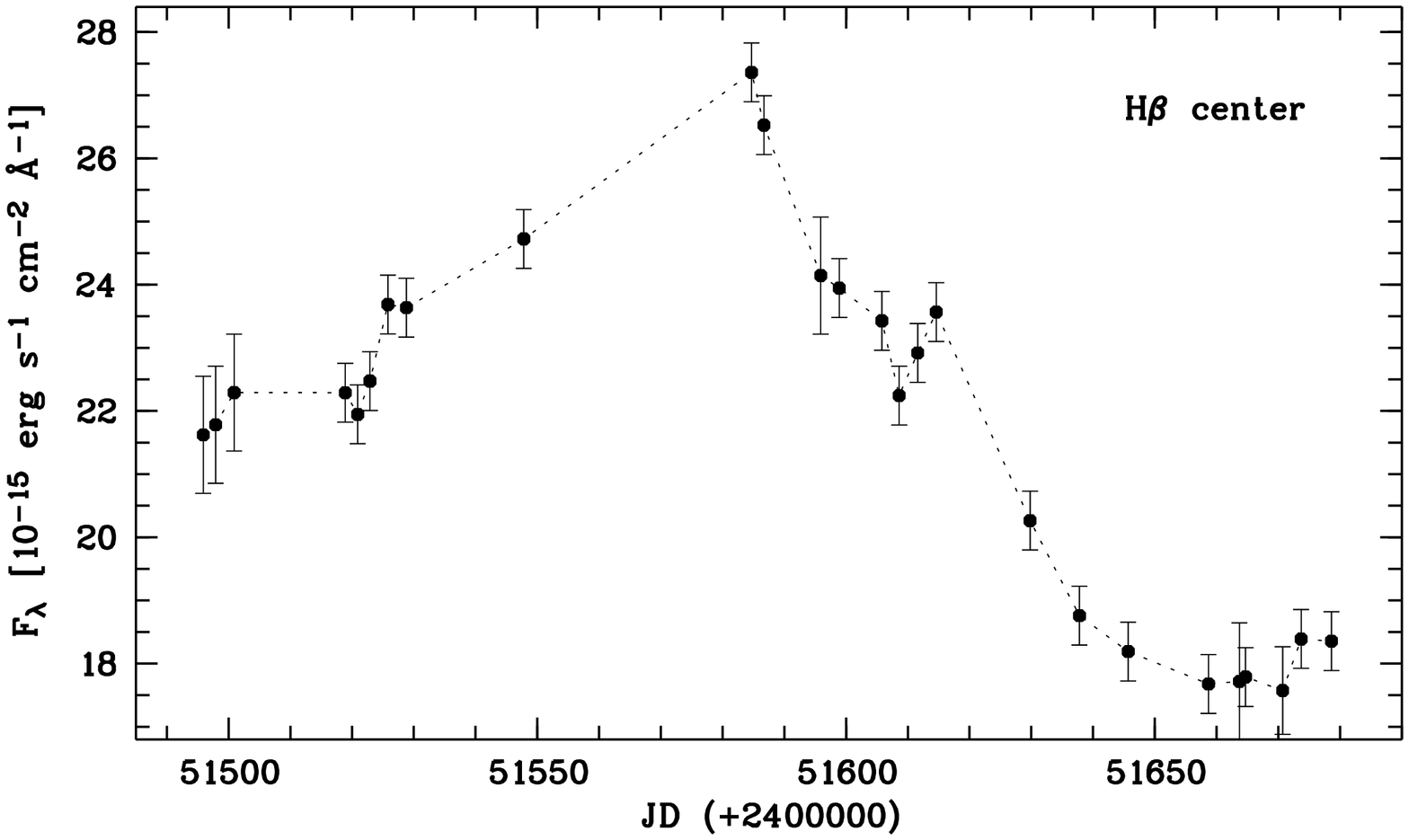}}
 \hbox{\includegraphics[bb=40 90 380 700,width=55mm,height=85mm,angle=270]
{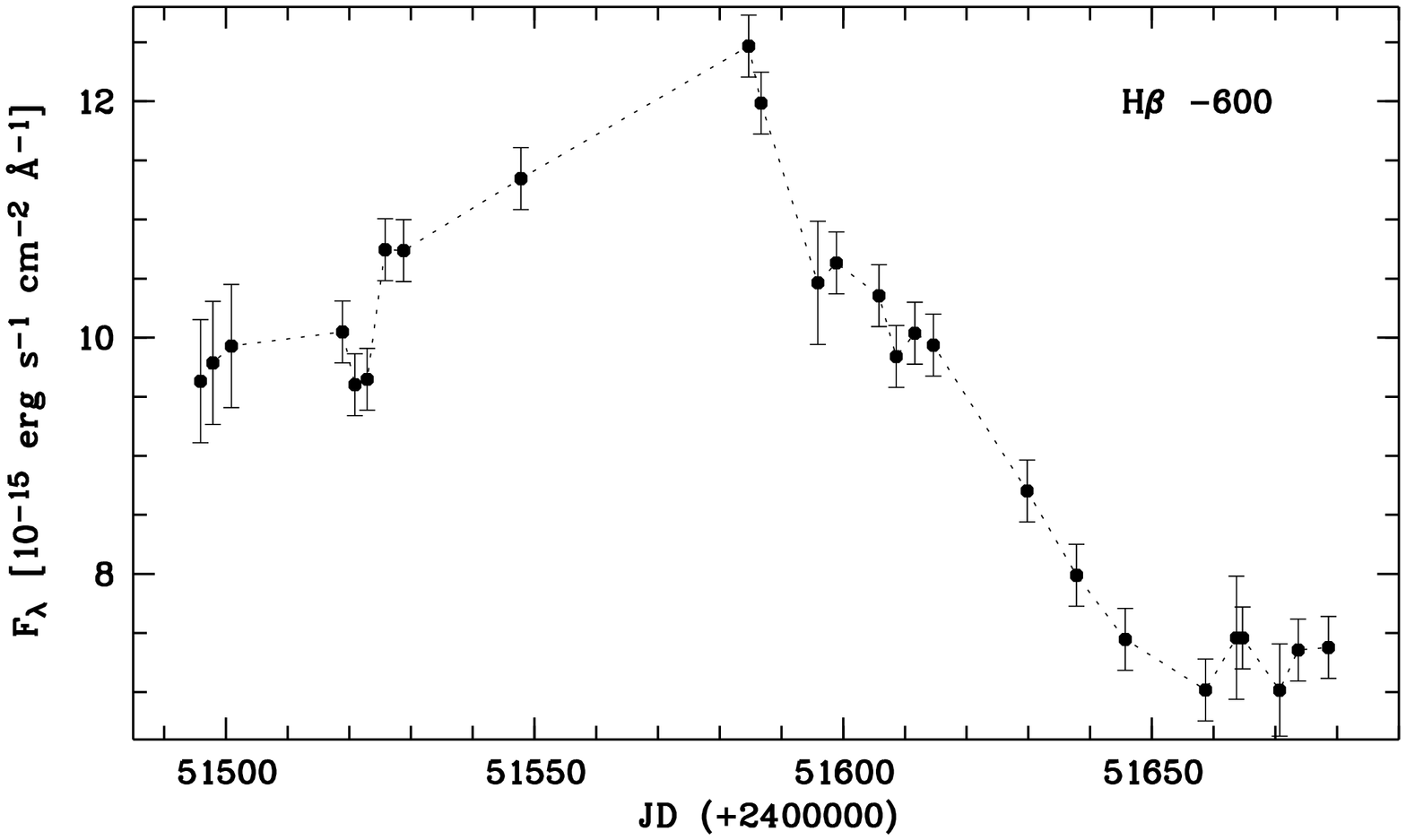}\hspace*{7mm}
       \includegraphics[bb=40 90 380 700,width=55mm,height=85mm,angle=270]
{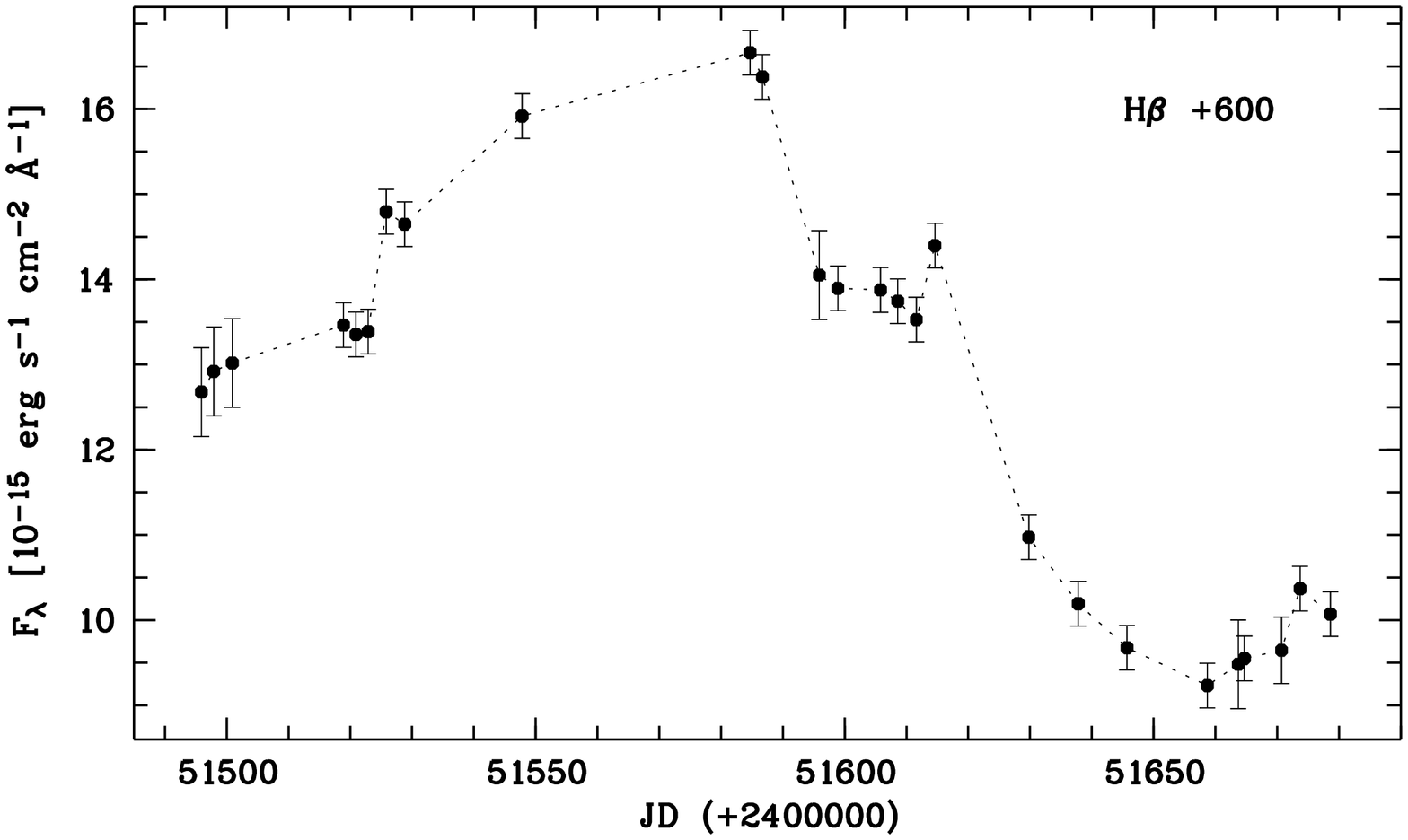}}
 \hbox{\includegraphics[bb=40 90 380 700,width=55mm,height=85mm,angle=270]
{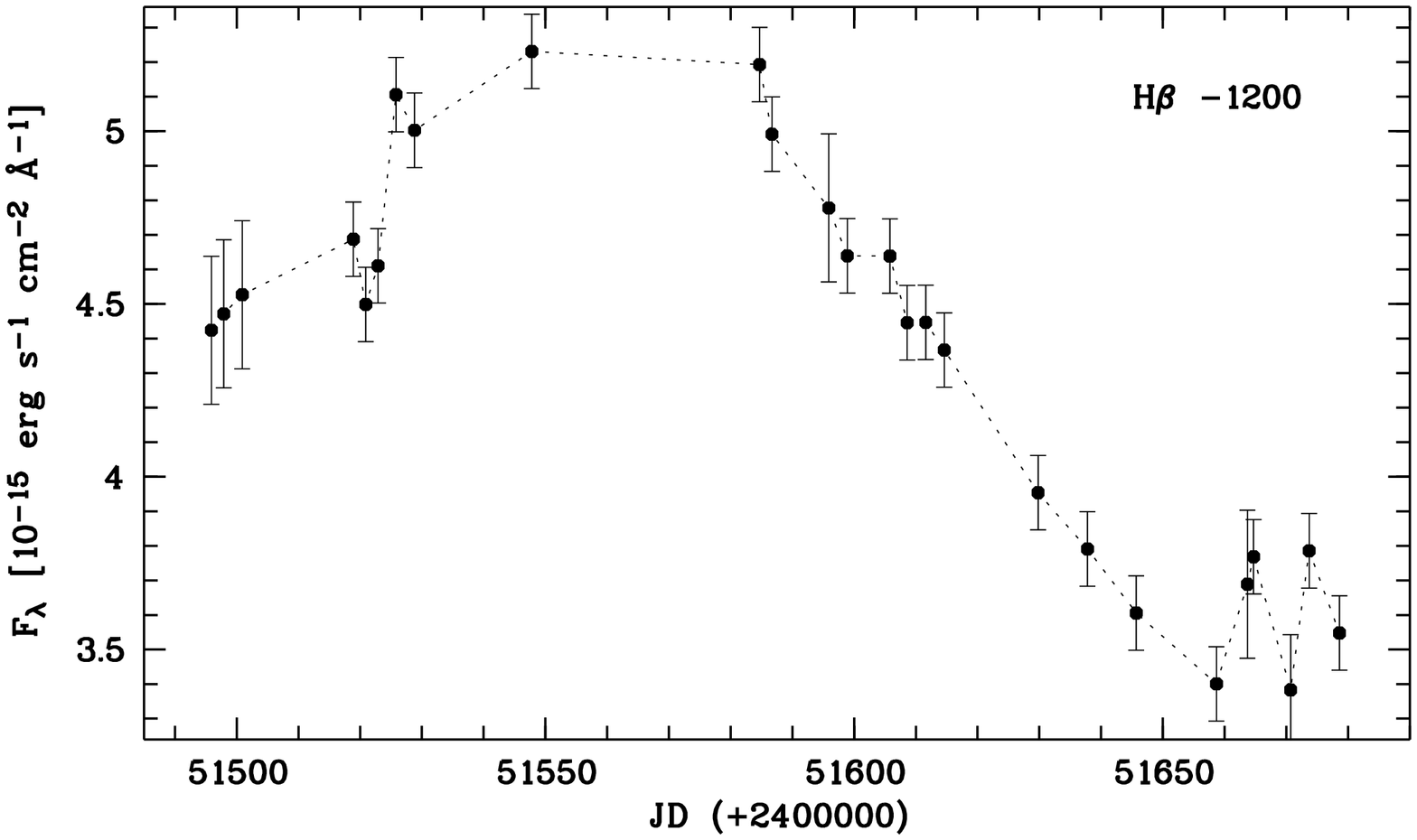}\hspace*{7mm}
       \includegraphics[bb=40 90 380 700,width=55mm,height=85mm,angle=270]
{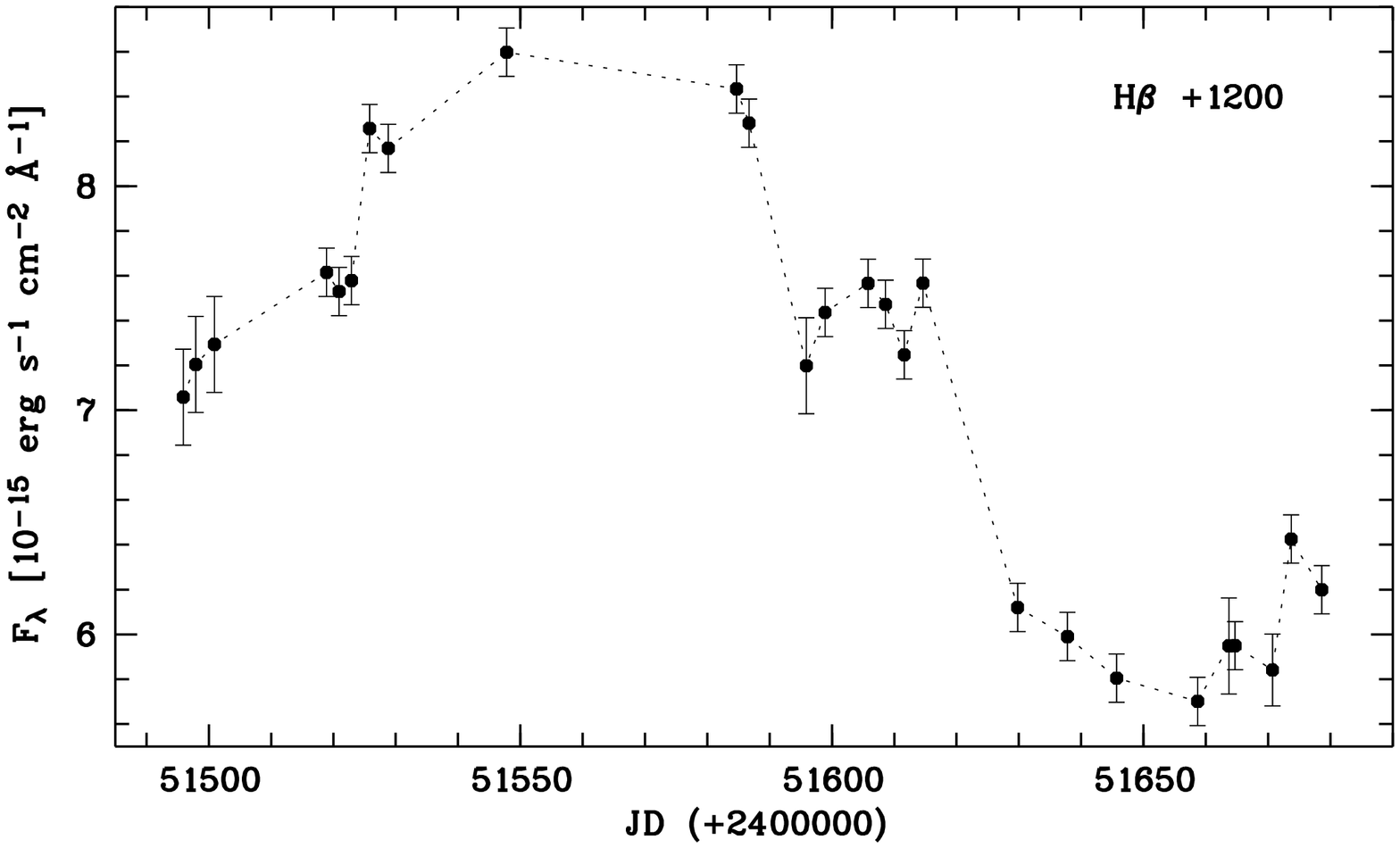}}
 \hbox{\includegraphics[bb=40 90 380 700,width=55mm,height=85mm,angle=270]
{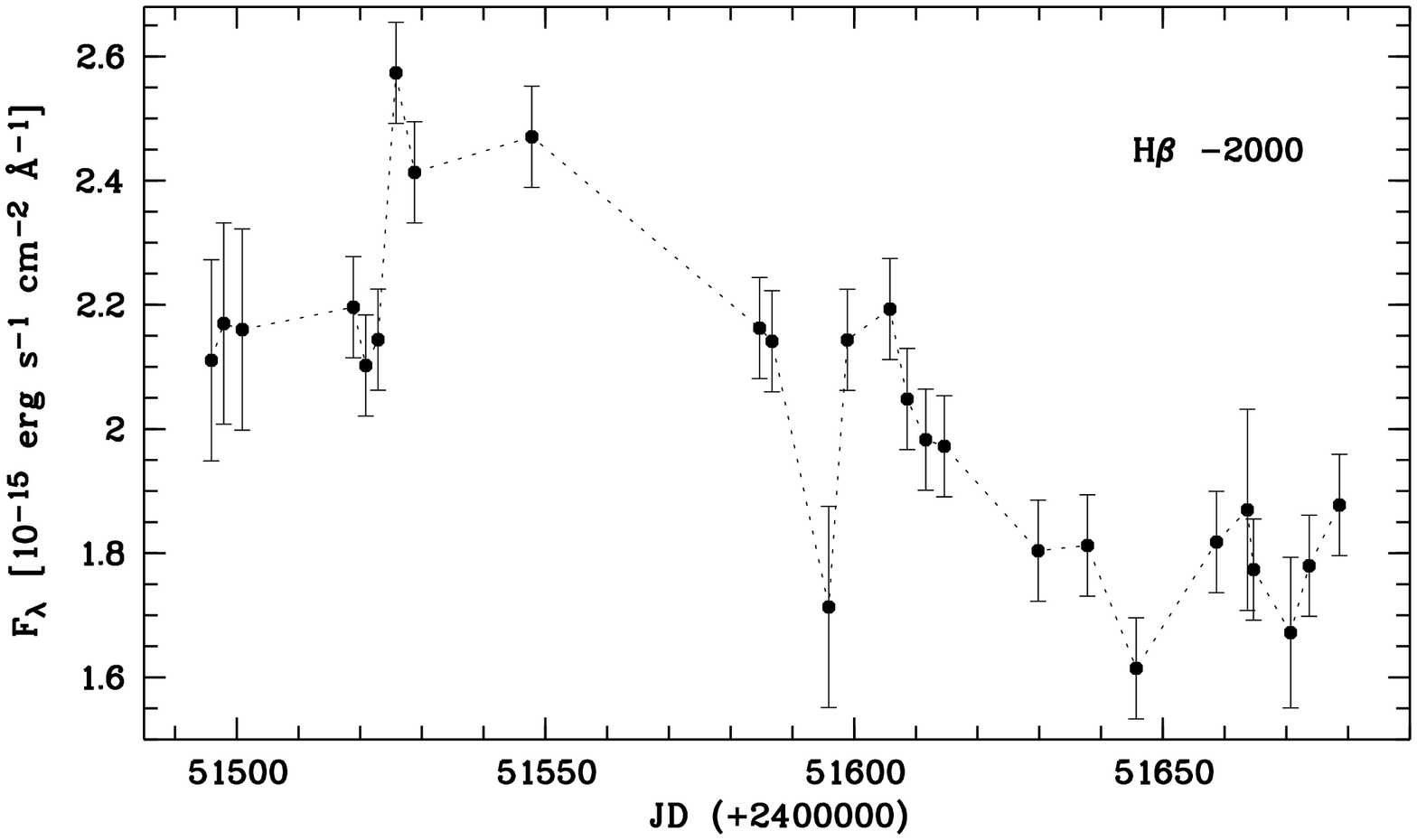}\hspace*{7mm}
       \includegraphics[bb=40 90 380 700,width=55mm,height=85mm,angle=270]
{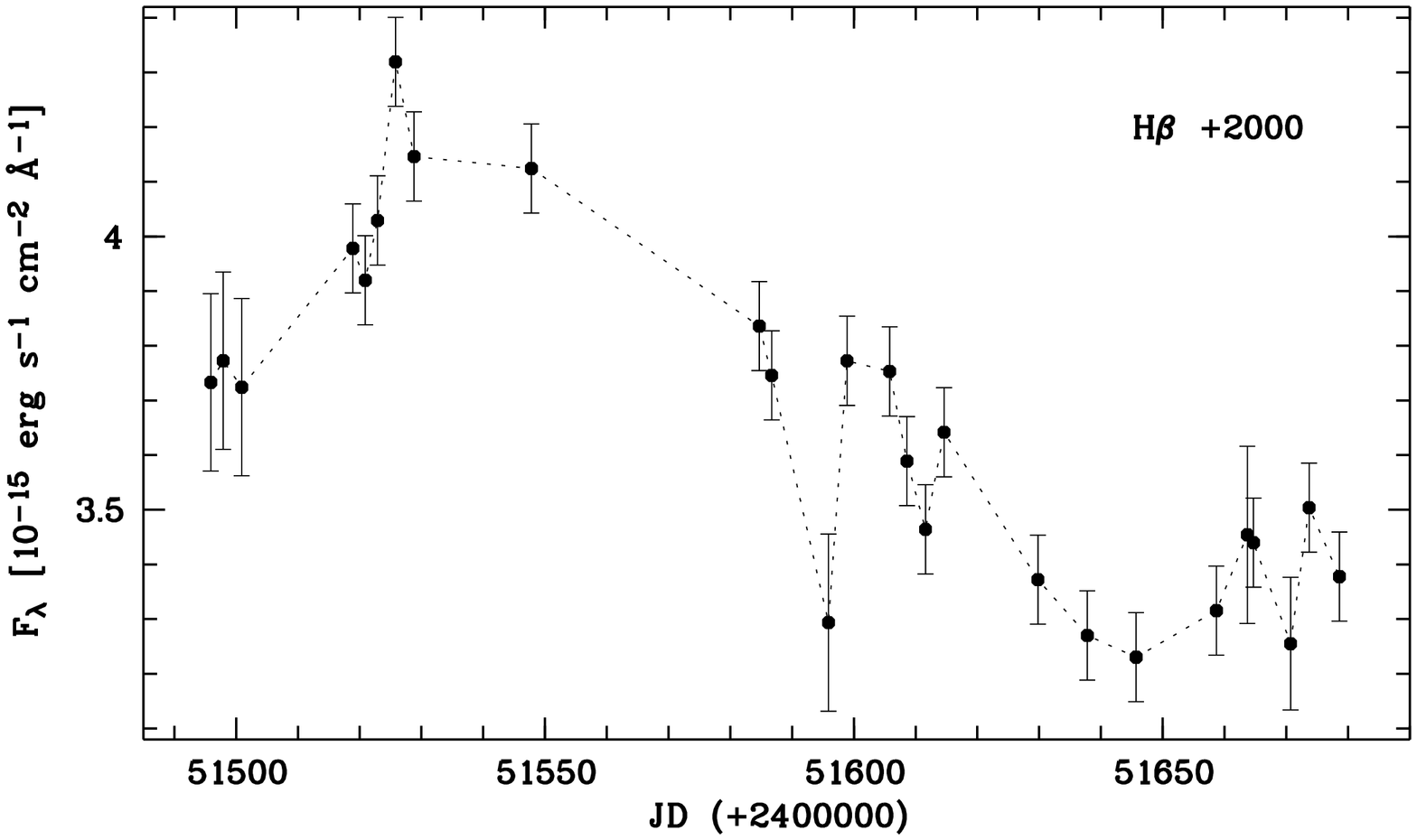}}
       \vspace*{5mm}
  \caption{Light curves of the continuum, of the H$\beta$ line center
 as well as of different blue and red H$\beta$ line wing segments
 ($v$ = $\pm$ 600, 1200, 2000 km/s, $\Delta v$ = 200 km/s) derived from our
 HET variability campaign of Mrk\,110.}
\end{figure*}
The light curves of individual line segments are noteworthy different. 
On the other hand light curves of the 
corresponding red and blue line segments differ less.

We computed cross-correlation functions (CCF) of all
H$\beta$ line segment light curves
with the 5100\AA\ continuum light curve (for details see Paper I).
The light curve of the central line segment
shows the largest delay.
The outer line wings segments
respond much faster to continuum variations than the inner ones.
We derived from the cross-correlation functions
a delay map of all H$\beta$ line segments ($\Delta v$ = 200 km/s)
which is shown in Fig.\ 3 in gray scale.
Contours of the correlation coefficient are overplotted.
\begin{figure}
\includegraphics[bb=37 167 576 625,width=88mm]{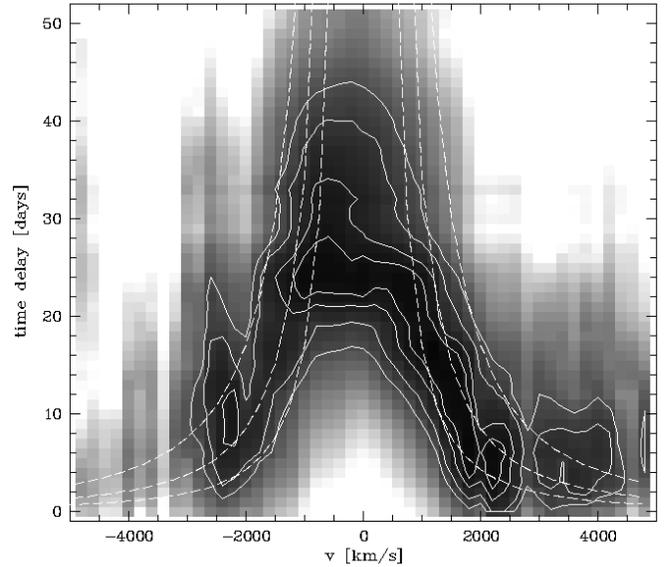}
\caption{
The 2-D CCF($\tau$,$v$) shows the correlation of the H$\beta$
line profile with continuum 
variations as a function of velocity and time delay (grey scale).
Contours of the correlation coefficient are overplotted at levels
of .85, .88, .91, .925 (solid lines).
The dashed curves show computed escape velocities for
central masses of 0.5, 1.0, 2.0 $\times\ 10^7 M_{\odot}$. (from bottom to top).}
\end{figure}
This 2-D CCF shows some clear trends. The
outer red and blue H$\beta$ line wings respond almost symmetrically
to continuum variations with a delay of about five days only.  
Towards the line center the delay
is getting systematically longer until up to about 30 days
at the line center.
Tests made with different H$\beta$
 velocity binning ($\Delta v$ = 100 -- 400 km/s)
gave qualitatively the same results.

The 2-D CCF($\tau$,$v$) is
mathematically very similar to a 2-D response function $\Psi$ (Welsh
 \cite{welsh01}).
Comparing our observed velocity-delay pattern with model calculations
(Perez et al. \cite{perez92}; Welsh \& Horne \cite{welsh91}; O'Brien et al.
 \cite{obrien94}) we can cut down
a lot of simple kinematical models of the BLR in Mrk\,110.
Both H$\beta$ line wings show the shortest delay with respect to the continuum
and react nearly simultaneously. Therefore we can rule out
radial inflow or outflow motions -- including biconical outflow.
Due to the fact that no short delays are observed at the center
one can furthermore rule out spherical BLR models
with chaotic virial velocity field or randomly oriented Keplerian orbits.
On the other hand, a Keplerian disk BLR model fulfills exactly 
the observed velocity-delay pattern:
the faster response of both line wings compared to the center.
There are further pieces of evidence from theoretical considerations
(Collin-Souffrin et al. \cite{collin88}, Emmering et al. \cite{emmering92})
 and from the
 observational side (Elvis \cite{elvis00}, Vestergaard et al.
 \cite{vestergaard00},
 Hutchings et al. \cite{hutchings01})
for a disk-like configuration of the broad-line region.

The 2-D CCF pattern follows the computed orbital velocities (Fig.\ 3). This demonstrates that the line emitting gas is gravitationally bound which is required for calculating a central black hole mass in AGN. 

A slightly faster and stronger response of the red line wing compared to the 
blue wing - as seen in Fig.\ 3 -
is predicted in the disk-wind model of the BLR of
Chiang \& Murray (\cite{chiang96}). These authors computed in detail
frequency-resolved response
functions of broad emission lines from the surface of an accretion
disk in the presence of a radiatively driven wind.
Their accretion disk-wind model matches
our observations of Mrk\,110 regarding to the blue-red-asymmetry.
The correlation of the red H$\beta$ wing with the continuum light curve 
is 5\% stronger than that of the blue
wing (Fig.\ 3). The blue H$\beta$ wing ($v$ = [--2000,--500] km/s) lags the
 red wing ($v$ = [+500,+2000] km/s) by
$2^{+2}_{-1}$ days. 

Furthermore, Murray \& Chiang (\cite{murray97}) demonstrated that a
 Keplerian disk with disk wind can
produce single-peaked broad emission lines as normally seen in most AGN.

\subsection{SMBH mass}

Black hole masses based on reverberation studies of integrated emission line
intensity variations 
 have been estimated for about three dozens
active galaxies (Wandel et al. \cite{wandel99}, Kaspi et al. \cite{kaspi00})
under the assumption that the emission line clouds
are gravitationally bound. 
The mean distance $R$ of the line emitting clouds and their velocity dispersion
$v$ derived from the mean width of the rms emission line profiles (FWHM)
are needed for computing a central black hole mass
(Koratkar \& Gaskell \cite{koratkar91}):
\[M = \frac{3}{2} v^{2} G^{-1} R .\]

We derived in Paper\,I
the central black hole mass in Mrk\,110
from the widths and delays of the integrated emission lines to
M= $1.6^{+0.3}_{-0.3}\times 10^{7} M_{\odot}$.
The velocity-delay map provides an improved method
to determine the central mass;
it is shown in Fig.\ 3  together with
computed escape velocities for central masses of 
0.5, 1.0, and 2.0 $\times\ 10^7 M_{\odot}$.
The escape velocities have been calculated with the formula given above where
$\tau$ is the CCF time delay and $R$=c$\tau$ the corresponding distance.
The observed delays of the resolved H$\beta$ line wings between
800 and 2000 km/s point to a central mass of
$1.0^{+1.0}_{-0.5}\times 10^7 M_{\odot}$ in Mrk\,110.
This black hole mass determination derived from the H$\beta$ line
profile matches that from the integrated line intensity variations
within the error limits and is an independent confirmation.

But, one has to keep in mind that in both cases
there are systematic uncertainties in the mass determination
(Krolik \cite{krolik01}). Most important, if the inclination
angle is unknown the derived mass is only a lower limit. 
The observed velocity-delay map does not drop down near the line 
center (Fig.\ 3) as expected from model 
calculations of edge-on disk models (Welsh \& Horne \cite{welsh91}; 
O'Brien et al. \cite{obrien94}).
This is a hint for a small inclination angle of the disk in Mrk\,110. 

\section{Summary and Outlook}

We demonstrated that
 the formation of the broad
H$\beta$ line emission in the wind of an accretion disk 
matches our observed 2-D variability pattern in Mrk\,110.
We expect the general picture will be confirmed
 in future variability campaigns on other galaxies that
broad emission lines in AGN are formed in the wind of an accretion disk.
Therefore, the difference between Seyfert 1 and Seyfert 2 galaxies
might be connected with the existence of such an accretion disk wind only.

\begin{acknowledgements}
      WK thanks the UT Astronomy Department for warm hospitality during
      his visit and W.F. Welsh and R. Robinson for many discussions 
      on this subject. 
      Part of this work has been supported by the
      \emph{Deut\-sche For\-schungs\-ge\-mein\-schaft, DFG\/} grant
      KO 857/24 and DAAD.
\end{acknowledgements}

\begin{thebibliography}{}

  \bibitem[1999]{bischoff99} Bischoff, K., \&  Kollatschny, W. 1999,
       A\&A, 345, 49

  \bibitem[1996]{chiang96} Chiang, J., \& Murray, N. 1996, ApJ, 466, 704

  \bibitem[1988]{collin88} Collin-Souffrin, S., Dyson, J.E., McDowell, J.C.,
      \& Perry J.J. 1988, MNRAS, 232, 539

  \bibitem[1996]{done96} Done, C., \& Krolik, J.H. 1996, ApJ, 463, 144

  \bibitem[2000]{elvis00} Elvis, M. 2000, ApJ, 545, 63

  \bibitem[1992]{emmering92} Emmering, R.T., Blandford, R.D., \& Shlosman, I.
       1992, ApJ, 385, 460

  \bibitem[2001]{hutchings01} Hutchings, J.B., Kriss, G.A., Green R.F. et al.
       2001, ApJ, 559, 173

  \bibitem[2000]{kaspi00} Kaspi, S., Smith, P.S., Netzer, H. et al.
       2000, ApJ, 533, 631

  \bibitem[1996]{kollatschny96} Kollatschny, W., \& Dietrich, M. 1996,
       A\&A, 314, 43

  \bibitem[1997]{kollatschny97} Kollatschny, W., \& Dietrich, M. 1997,
       A\&A, 323, 5

  \bibitem[2001]{kollatschny01} Kollatschny, W., Bischoff, K., Robinson, E.L.,
      Welsh, W.F., \& Hill, G.J.  2001, A\&A, 379, 125 (Paper\,I)

  \bibitem[1991]{koratkar91} Koratkar, A., \& Gaskell, M.  1991, ApJ, 370, L61

  \bibitem[2001]{krolik01} Krolik, J.\ H.\ 2001, ApJ, 551, 72

  \bibitem[1997]{murray97} Murray, N., \& Chiang, J.  1997, ApJ, 474, 91

  \bibitem[1994]{obrien94} O'Brien, P.T., Goad, M.R., \& Gondhalekar, P.M.,
         1994, MNRAS, 268, 845

  \bibitem[1992]{perez92} Perez, E., Robinson, A., \& de la Fuente, L.
         1992, MNRAS, 256, 103

  \bibitem[1998]{peterson98} Peterson, B.M., Wanders, I., Bertram, R., et al.
       1998, ApJ, 501, 82

  \bibitem[1996]{ulrich96} Ulrich, M.-H., \& Horne, K. 1996, MNRAS, 283, 748

  \bibitem[2000]{vestergaard00} Vestergaard, M., Wilkes, B.J., \& Barthel, P.D.
       2000, ApJ, 538, L103

  \bibitem[1999]{wandel99} Wandel, A., Peterson, B.M., \& Malkan, M.A.
       1999, ApJ, 526, 579

  \bibitem[1991]{welsh91} Welsh, W.F., \& Horne, K.  1991, ApJ, 379, 586

  \bibitem[2001]{welsh01} Welsh, W.F. 2001, in: Peterson et al. (eds.): 
       Probing the Physics of AGN, ASP Conf.Ser. 224, p.123

\end{thebibliography}
\end{document}